\documentclass[
]{ceurart}

\usepackage{listings}

\usepackage{xcolor}
\usepackage{todonotes}

\begin{document}

\copyrightyear{2022}
\copyrightclause{Copyright for this paper by its authors.
  Use permitted under Creative Commons License Attribution 4.0
  International (CC BY 4.0).}

\conference{KCAP'25: The Thirteenth Int;ernational Conference on Knowledge Capture,
  December 10--12, 2025, Dayton, OH}

\title{LIT-GRAPH: Evaluating Deep vs. Shallow Graph Embeddings for High-Quality Text Recommendation in Domain-Specific Knowledge Graphs}

% \tnotetext[1]{You can use this document as the template for preparing your
%   publication. We recommend using the latest version of the ceurart style.}

\author[1]{Nirmal Gelal}[%
email=ngelal@ksu.edu,
]
\address[1]{Department of Computer Science, Kansas State University}

\author[2]{Chloe Snow}[%
email=chlo24@ksu.edu,
]
\address[2]{Department of Curriculum and Instruction, Kansas State University}

\author[1]{Kathleen M. Jagodnik}[%
email=kmjagodnik@ksu.edu,
]

\author[2]{Ambyr Rios}[%
email=ambyrrios@ksu.edu,
]

\author[1]{Hande Küçük McGinty}[%
email=hande@ksu.edu,
]
\cormark[1]

\cortext[1]{Corresponding author.}
% \fntext[1]{These authors contributed equally.}

\begin{abstract}
  This study presents LIT-GRAPH (Literature Graph for Recommendation and Pedagogical Heuristics), a novel knowledge graph-based recommendation system designed to scaffold high school English teachers in selecting diverse, pedagogically aligned instructional literature. The system is built upon an ontology for English literature, addressing the challenge of curriculum stagnation, where we compare four graph embedding paradigms: DeepWalk, Biased Random Walk (BRW), Hybrid (concatenated DeepWalk and BRW vectors), and the deep model Relational Graph Convolutional Network (R-GCN). Results reveal a critical divergence: while shallow models excelled in structural link prediction, R-GCN dominated semantic ranking. By leveraging relation-specific message passing, the deep model prioritizes pedagogical relevance over raw connectivity, resulting in superior, high-quality, domain-specific recommendations.
\end{abstract}


\maketitle

\section{Introduction}
The selection of diverse, thematically aligned instructional literature remains a significant hurdle for high school English teachers, due to time constraints and limited access to resources. This often compels educators to rely on a stagnant canon of familiar texts, resulting in a curriculum that may lack diversity and fail to resonate with students' identities and interests. To address this, T-TExTS \cite{gelal2025t} introduced an ontology-driven recommendation framework tailored to the specific pedagogical needs of English Language Arts (ELA) instruction. This framework utilizes a Knowledge Graph (KG) enriched with expert-curated metadata, including Lexile levels, themes, and literary elements.

This paper serves as an extension of that foundational work, specifically investigating the representational learning component of the system. A core challenge in applying KGs to this domain is determining the optimal embedding method to power the recommendation engine. We compare shallow (transductive) models against deep (inductive) models to determine which approach best captures the rich semantics of a small, high-density educational graph. We hypothesize that deep embedding methods will outperform shallow methods in generating recommendation-friendly representations.

\section{Methodology}
The dataset used in this study is identical to the one employed in our previous work, in which we collected 98 English Literature texts. Data curation was performed through consultation with domain experts, secondary ELA teachers, and Kansas State University faculty, alongside an analysis of Kansas high school text selection.

We utilized the Knowledge Acquisition and Representation Methodology (KNARM)\cite{knarm} to construct a novel ontology. In contrast to our previous high expressivity Web Ontology Language (OWL) implementation, this study adopts a lightweight schema optimized for direct Resource Description Framework (RDF) serialization that aligns the conceptual model directly with the underlying triple-store architecture, eliminating the overhead of stripping formal restrictions (e.g., owl:Restrictions, owl:onProperty). The resulting KG comprises 364 classes, 11 object properties, 6 data properties, 3,303 triples, and 568 unique entities.

\subsection{Embedding models}
\begin{figure}[htbp]
  \centering
  \includegraphics[width=0.8\textwidth]{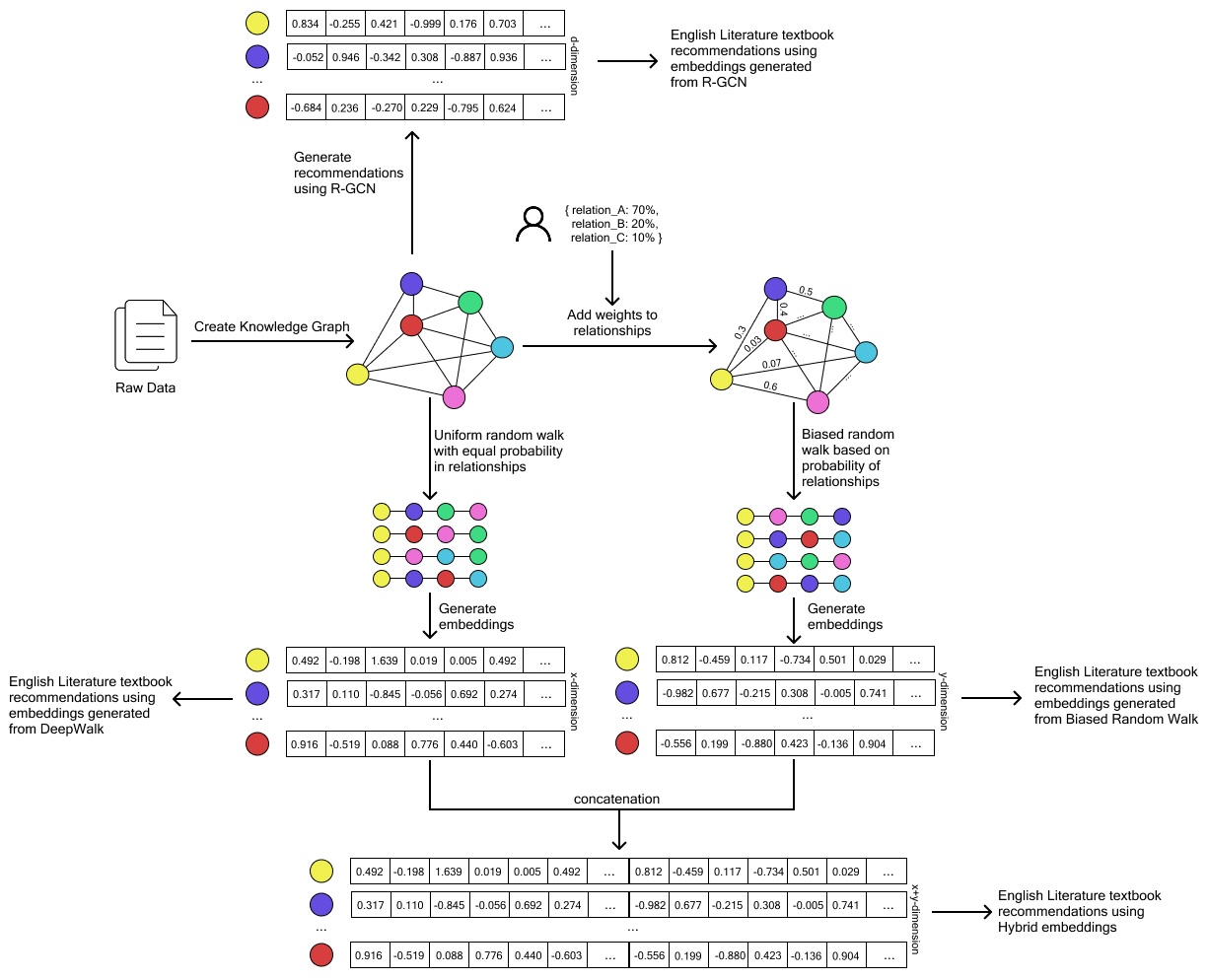}
  \caption{Workflow of the English Literature Text Recommender System.\\
  \footnotesize{Note: The Hybrid model is the concatenation of vector outputs from DeepWalk (DW) and Biased Random Walk (BRW). R-GCN: Relational Graph Convolutional Network.}
  }
  \label{fig:workflow}
\end{figure}

We implemented four embedding strategies to convert KG entities into vector representations. The complete workflow of the system, from raw data processing to final recommendation generation, is illustrated in Figure~\ref{fig:workflow}.

\subsubsection{Shallow Models}

The embedding techniques in this study are classified as shallow models, which learn latent representations by mapping nodes directly to a lookup table rather than through multi-layered neural architectures. This approach prioritizes computational efficiency and local structural preservation, providing a robust baseline for capturing both the topological and expert-defined features of the Knowledge Graph.

We implemented three variations of this paradigm: \textbf{DeepWalk} \cite{perozzi2014deepwalk}, which utilizes random walks, and Skip-gram architecture to capture unbiased structural proximity; a \textbf{Biased Random Walk}\cite[cf.][]{cochez2017biased}, which replaces uniform sampling with an expert-guided weighting system that prioritizes semantically rich relationships like \textit{hasTheme}; and a \textbf{Hybrid} technique. The Hybrid approach concatenates these vectors to synthesize a dual signal, balancing raw network topology with specialized, expert-defined semantic hierarchies to maximize predictive power.

\subsubsection{Deep Models}
To address the inherent limitations of shallow, transductive methods, this study incorporates a deep relational model to capture complex, multi-hop dependencies. Unlike static lookup tables, our approach utilizes an inductive learning paradigm built upon message-passing mechanisms, where node representations are generated by iteratively aggregating feature vectors from local neighborhoods. This allows the model to learn the underlying functions governing the graph's global topology rather than simply memorizing local proximity.

Specifically, we implemented the \textbf{Relational Graph Convolutional Network (R-GCN)} \cite{schlichtkrull2018modeling} to manage the multi-relational complexity of the Knowledge Graph. By employing relation-specific weight matrices, the R-GCN differentiates incoming signals based on their relationship type, preserving the unique semantic influence of pedagogical links versus auxiliary metadata. This architecture enables the generation of expressive embeddings utilizing the full ontological schema.

\section{Experimental Setup}
The experimental workflow was designed to first optimize node representations through a link prediction task before evaluating their utility for recommendation. The Knowledge Graph edges were split into training (80\%), validation (10\%), and testing (10\%) sets. To employ high-quality embeddings, we utilized link prediction as an intermediate evaluative step, employing negative sampling to generate false triples by randomly corrupting head or tail entities, ensuring no overlap with valid triples. During this phase, the models were trained to maximize the Area Under the Curve (AUC), optimizing the latent space to effectively discriminate between true and false triples. This rigorous objective ensures that the resulting embeddings accurately reflect the underlying relational structure of the graph.

Once the representations were tuned, we utilized the optimized embeddings to measure recommendation quality through a candidate ranking task. Using cosine similarity, we ranked potential books against target entities to evaluate how well the learned features translated to user-relevant suggestions. These generated rankings were then evaluated against a ground truth dataset manually curated by domain experts. The quality of these recommendations was quantified using three standard ranking metrics: Hits@K, Mean Reciprocal Rank (MRR), and Normalized Discounted Cumulative Gain (nDCG). To ensure a robust comparison across both shallow and deep architectures, all hyperparameters, including walk parameters for shallow models and hidden dimensions or weight decay for the R-GCN model, were systematically optimized using Optuna\footnote{\url{https://github.com/optuna/optuna}}.

\section{Results, Discussion, and Limitations}
The evaluation of our embedding models reveals a distinct divergence between link prediction accuracy and recommendation ranking performance. As shown in Table~\ref{tab:results}, the shallow DeepWalk model achieved the highest AUC (0.9737), significantly outperforming the deep R-GCN model (0.7405). This suggests that uniform random walks are highly effective at capturing the global topological structure required for binary link classification. However, this high link prediction accuracy did not directly translate to superior recommendation quality.

When evaluating the models on recommendation ranking metrics, the deep architecture demonstrated a clear advantage. As detailed in Table~\ref{tab:results}, the R-GCN model consistently outperformed shallow counterparts across most ranking metrics, achieving the highest Hits@10 (0.7368) and nDCG@10 (0.4985). This supports the hypothesis that deep embedding approaches, specifically those utilizing relational message-passing, are better suited for recommendation in knowledge-augmented datasets. The R-GCN's ability to model distinct relation types allows it to capture a richer semantic context optimized for item ordering, even when its binary discrimination capability, measured by AUC, is lower than that of sampling-based methods. Notably, the Hybrid model remained competitive at the Hits@5 threshold (0.6842), suggesting that expert-guided biased walks can effectively augment local neighborhood retrieval.

\begin{table*}[t]
\centering
\caption{Link Prediction (AUC) and Recommendation Ranking Performance across models}
\label{tab:results}
\begin{tabular}{l c c c c c}
\toprule
\textbf{Model} & \textbf{AUC $\uparrow$} & \textbf{Hits@10 $\uparrow$} & \textbf{Hits@5 $\uparrow$} & \textbf{MRR $\uparrow$} & \textbf{nDCG@10 $\uparrow$} \\
\midrule
DeepWalk        & \textbf{0.9737} & 0.6316 & 0.6316 & 0.4264 & 0.4549 \\
Biased RW       & 0.7805          & 0.6316 & 0.6316 & 0.4035 & 0.4319 \\
Hybrid          & 0.8091          & 0.6842 & \textbf{0.6842} & 0.4185 & 0.4550 \\
R-GCN           & 0.7405          & \textbf{0.7368} & 0.5789 & \textbf{0.4449} & \textbf{0.4985} \\
\bottomrule
\end{tabular}
\end{table*}

Limitations primarily stem from the dataset size (98 books), constrained by the scarcity of datasets with specialized pedagogical metadata. While expert curation ensured high fidelity, the resulting low graph volume may limit the generalizability of the trained models. Furthermore, standard metrics like AUC and Hits@K, optimized for large-scale contexts, may not fully capture the practical pedagogical utility of recommendations within such small, high-density, expert-driven domains.

\section{Conclusion and Future Work}
This study extended the evaluation of LIT-GRAPH to assess the impact of embedding architectures. Results reveal a significant divergence between metrics: while shallow models like DeepWalk excelled in structural link prediction, the deep R-GCN model proved superior for semantic ranking. By leveraging relation-specific message passing, R-GCN effectively utilized the schema to prioritize pedagogical relevance over raw connectivity, validating the necessity of deep learning for domain-specific educational recommendations.

Future work focuses on scaling the dataset by leveraging LLMs for automated KG construction \cite{dalal2026echo} and semi-automated annotation to mitigate curation bottlenecks. Additionally, we aim to develop specialized metrics that capture pedagogical alignment beyond traditional ranking scores. Finally, integrating Graph Retrieval-Augmented Generation (GraphRAG) could evolve the system into an interactive, explanatory assistant, significantly enhancing its transparency and utility for educators.

\bibliography{main}

\end{document}